\documentclass[a4paper, 11pt] {article}
\usepackage{latexsym}
\usepackage{graphicx}
\usepackage{epsfig}

\begin{document}
\newcommand{\eg}{{\it e.g.}}
\newcommand{\etal}{{\it et. al.}}
\newcommand{\ie}{{\it i.e.}}
\newcommand{\be}{\begin{equation}}
\newcommand{\dd}{\displaystyle}
\newcommand{\ee}{\end{equation}}
\newcommand{\bea}{\begin{eqnarray}}
\newcommand{\eea}{\end{eqnarray}}
\newcommand{\bef}{\begin{figure}}
\newcommand{\eef}{\end{figure}}
\newcommand{\bce}{\begin{center}}
\newcommand{\ece}{\end{center}}
\def\lsim{\mathrel{\rlap{\lower4pt\hbox{\hskip1pt$\sim$}}
    \raise1pt\hbox{$<$}}}         
\def\gsim{\mathrel{\rlap{\lower4pt\hbox{\hskip1pt$\sim$}}
    \raise1pt\hbox{$>$}}}         

\begin{center}
{\Large Do quantum effects hold together DNA condensates?}

\vskip 1cm

Alfredo Iorio$^{(a)}$, Samik Sen$^{(b)}$, Siddhartha Sen$^{(c,d)}$ \\
$^{(a)}$ Institute of Particle and Nuclear Physics, Charles University of Prague \\
V Hole\v{s}ovickach 2, 180 00 Prague 8 - Czech Republic \\
$^{(b)}$  Rocwood 27, Stillorgan (Co. Dublin) - Ireland \\
$^{(c)}$ School of Mathematical Sciences, University College Dublin \\
Belfield, Dublin 4 - Ireland \\
$^{(d)}$ Indian Association for the Cultivation of Science \\
Jadavpur, Calcutta 700032 - India

\vskip 0.5cm
E-mails: iorio@ipnp.troja.mff.cuni.cz, samiksen@gmail.com, sen@maths.ucd.ie

\vskip 1cm

\today

\end{center}

\begin{abstract}
The classical electrostatic interaction between DNA molecules in water in the presence of
counterions is reconsidered and we propose it is governed by a modified
Poisson-Boltzmann equation. Quantum fluctuations are then studied and shown to
lead to a vacuum interaction that is numerically computed for several configurations
of many DNA strands and found to be strongly many-body. This Casimir vacuum interaction can be
the ``glue'' holding together DNA molecules into aggregates.

\end{abstract}

\vskip 0.5cm



\vskip 0.5cm

The theoretical study of the formation and stability of DNA aggregates achieved an
important understanding of these phenomena that, nonetheless, still appear mysterious in
many respects \cite{gelbart2000}. A DNA molecule in aqueous solution ionizes and gives rise to a highly
charged anion that, in the presence of cations, binds them by the Oosawa-Manning (OM)
condensation \cite{oosawa, manning} (for a review see, e.g., \cite{gelbart2000, Levin:2002gj, nguyen1}).
When about 90 per cent of the DNA negative charge is screened and when the cations have a
specific valency $+ k$ (usually $k = 3$ or $k = 4$) the DNA strands collapse to form rod-like,
spheroidal and toroidal aggregates \cite{Hud}.

There are two approaches followed to understand these features: In one the detailed charge distribution
of the DNA is used to calculate the electrostatic interaction between two DNA strands using a
{\it linearized} Poisson-Boltzmann (PB) (or Debye-H\"{u}ckel (DH)) equation
\cite{kornyshev1999} (see also \cite{KLreview}). Within this approach the attraction between
two like-sign charged but suitably oriented DNA helical strands and the specificity of cations driving the attraction
can be predicted. However, it was also realized that
an assembly of strands in an hexagonal array is a
frustrated-charge system \cite{harreis2003}. Thus the reasons why within this approach
the DNA aggregates should form and be stable is not clear, unless other
forces are invoked.

In alternative approaches the surface of a single DNA strand is
treated as a two-dimensional complex system and statistical
mechanical arguments lead to the counterion-mediated attraction
between DNA strands \cite{oosawabook, groenbech, Ha1997, parsegian1998}. The
key idea there is that the condensation is triggered by local
correlations and thermal fluctuations not present in the mean-field
PB approach. The electrostatic interaction again plays an essential role and an
assembly of DNA strands forming an hexagonal bundle is found to be
a frustrated-charge system \cite{bruinsmapre}. Here, again, the formation and stability
of aggregates is not clear if other forces are not present.

In this letter we propose that quantum vacuum fluctuations are likely to be the
solution to some of these puzzles. In particular we propose that the Casimir interaction due to these
fluctuations is the ``glue'' that holds DNA aggregates together.

The PB equation we need to consider is
\begin{equation}\label{PoissonBoltzmann}
    \nabla^2 \Phi (\vec{x}) = - \frac{4 \pi}{\epsilon} \rho(\vec{x}, T) = \frac{8 \pi}{\epsilon} k e n_0 \sinh \left( \frac{k e \Phi(\vec{x})}{k_B T} \right) \;,
\end{equation}
where  $\Phi(\vec{x})$ is the electrostatic potential due to the DNA strand (seen as a negatively charged rod immersed
in water at room temperature, $T \simeq 300$K, with dissolved salt whose ions have valency $z = \pm k$, with $k = 1, 2, ...$) and to
the ions, the medium has dielectric constant $\epsilon$, and the charge distribution of the composite system DNA-salt is
$\rho(\vec{x}, T) = \rho_{\rm DNA} (\vec{x}, T) + k e (n_+
(\vec{x}, T) - n_- (\vec{x}, T))$, with $n_\pm (\vec{x}, T)$ the concentration (density) of ions following
a Boltzmann distribution. Usually $\rho_{\rm DNA}$ is not included into the
PB equation (\ref{PoissonBoltzmann}). We demand, instead, that it obeys a Boltzmann distribution
law as for the ions \cite{iss}
\begin{equation}
\rho_{\rm DNA} (\vec{x}, T) = - n^0_{\rm DNA} (\vec{x}) |q| \exp
\left(  \frac{|q| \Phi (\vec{x})}{k_B T} \right)
\end{equation}
where $q < 0$ is the charge of the DNA strand with $n^0_{\rm DNA} (\vec{x}) = \sum_{i=1}^N \nu_i (z_i) \delta^2 (\vec{x}_\bot - \vec{l}_i)$ and $N$ the number if strands.
This charge density function also defines our approximations: we model the DNA strands
as infinite lines all parallel to the $z$-axis and located at
$\vec{l}_i$ in the $x-y$ plane with the coefficients $\nu_i (z_i)$
carrying information on the charge structure of the DNA strand. We
further simplify our model by taking $\nu_i (z_i) = \nu =$~constant,
$\forall i = 1, ..., N$.

Our concern is to study the interaction among DNA strands {\it
after} at least 90 per cent of the negative charge has been
screened via the OM condensation \cite{oosawa}, \cite{manning}, as this is the reported critical
value for collapse. It is then reasonable to consider $\Phi$ small. Thus, expanding
the exponentials till the first order we obtain
\begin{equation}\label{DHmod}
\left[ - \partial^2_z - \nabla_\bot^2 + \mu^2 + \lambda \sum_{i=1}^N \delta^{(2)} (\vec{x}_\bot - \vec{l}_i) \right] \Phi(\vec{x}) = J \;,
\end{equation}
which is a {\it modified} DH equation. Here $\mu^2 = k^2 \kappa^2$, with $\kappa^{-1} = (\epsilon k_B T/(8 \pi e^2 n_0))^{1/2}$ the Debye screening length, $\lambda = 4 \pi \nu |q|^2 / \epsilon k_B T$, and $J =  - (1 / \epsilon) 4 \pi |q| \nu \sum_{i=1}^N \delta^{(2)} (\vec{x}_\bot - \vec{l}_i)$.

\begin{figure}
 \centering
  \includegraphics[height=.2\textheight]{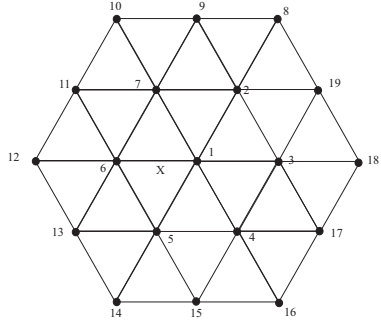}
  \caption{The configuration used for 19 DNA strands where $x$ is the distance between
  nearest neighbors. }
\label{steps}
\end{figure}

We now consider small time-dependent fluctuations, $\Phi
(\vec{x}) \to \Phi (\vec{x}) + \phi (\vec{x}, t)$, where $\Phi$
satisfies Eq.~(\ref{DHmod}) that descends from the action
\begin{equation}\label{actionPhi}
{\cal A} (\Phi) = \int d^4 x  \left( \frac{1}{2} \Phi [- \partial^2_z - \nabla_\bot^2 + \mu^2 + \lambda \sum_{i=1}^N \delta^{(2)} (\vec{x}_\bot - \vec{l}_i)] \Phi + J \Phi \right) \;,
\end{equation}
where we use units $\hbar = c = 1$, with $c$ the velocity of light in the medium and, for the sake of clarity,
we included an integration over time $\int_0^\tau dt$ even though the functions are
time-independent. We then demand that to the fluctuation field $\phi$ as well is associated an action that is a suitable
modification of (\ref{actionPhi}), namely
\begin{equation}
\bar{{\cal A}} (\phi) = \int d^4 x \frac{1}{2} \phi \left( - \partial_t^2 - \partial^2_z - \nabla_\bot^2 + \mu^2 + \lambda \sum_{i=1}^N \delta^{(2)} (\vec{x}_\bot - \vec{l}_i) \right) \phi \;.
\end{equation}
Note that in $\bar{{\cal A}} (\phi)$ the term with the coupling to the ``external current'' $J$ is zero because
$\int d^4 x J \phi = \int d^3 x J \int_0^\tau dt \phi = 0$ as required for fluctuating
fields. The way to consider the effects of $\phi$ is to average these fluctuations
out to obtain an effective action ${\cal A}_{\rm eff} (\Phi)$. This is done by considering the
generating functional $Z[\Phi, \phi]  =  \int [D\Phi] e^{i {\cal A} (\Phi)} \int [D\phi] e^{i \bar{{\cal A}} (\phi)}
= \int [D\Phi] e^{- ( {\cal A} (\Phi) + {\rm corrections})}$ where we Wick rotate on the time direction $t \to i t$, and
identify ${\cal A}_{\rm eff} (\Phi) = {\cal A} (\Phi) +$
corrections. The result is \cite{iss} ${\cal A}_{\rm eff} (\Phi) = {\cal A} (\Phi)
+ {\cal E} \tau$, with ${\cal E} = \frac{1}{2} \int_{-\infty}^{+\infty} \frac{d p}{2 \pi} \int_0^{+\infty} dE \rho (E) \sqrt{E + p^2}$, where
$\rho(E)$ is the density of states.

\begin{figure}
 \centering
  \includegraphics[height=.3\textheight]{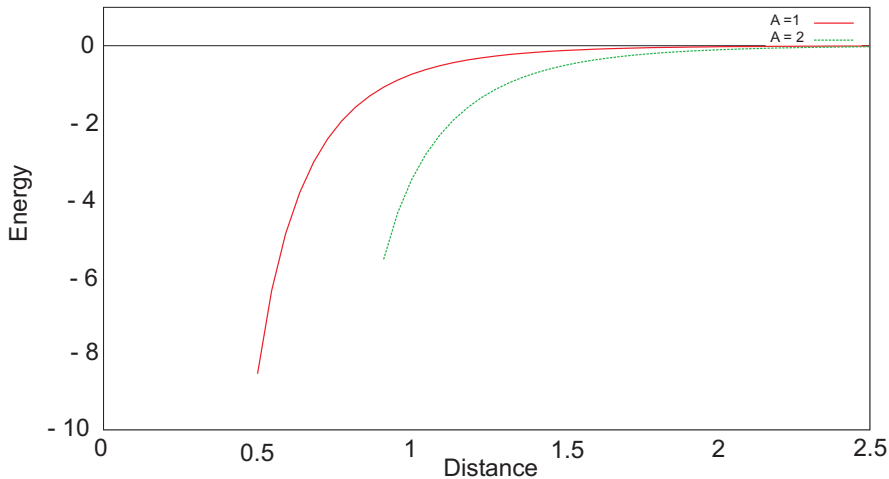}
  \caption{Energy of interaction of two DNA strands. The lower (upper) curve corresponds to $a =2$ ($a=1$). Distances
  are measured in units of $1/ \mu \sim$ ${\cal O} (10)$ \AA, while energy units are estimated to be ${\cal O} (10^{-1}) - {\cal O} (1)$~eV.}
\label{2_String_Energy}
\end{figure}

This energy is of the form ${\cal E} = (1/2) \sum \omega$, i.e. it is the zero point Casimir energy of the system and it can be determined \cite{iss}, \cite{Scardicchio:2005hh} (see also \cite{Jaffe:2005wg}) and is
\begin{equation}\label{Elndet}
{\cal E} = \frac{\hbar c}{8 \pi} \int_0^\infty dE \ln \left[ \det \left( \delta_{i j} -  \frac{K_0 (\sqrt{E
+ \mu^2} \; l_{i j})}{\ln(\sqrt{E + \mu^2} / M)} (1 - \delta_{i j})
\right) \right] \;,
\end{equation}
where we reintroduced $\hbar$ and $c$, $K_0 (x)$ is the modified Bessel function of the second kind
of order zero, $\mu$ is the mass (inverse length) scale parameter
introduced earlier, $M$ is a further mass (inverse length) scale
parameter that satisfies $M < \mu$ and $l_{i j} = |\vec{l}_i -
\vec{l}_j|$ are the relative distances between DNA strands. Note
that $\lambda$, the other parameter of the theory, is first
renormalized $\lambda \to \lambda_R$ and then adsorbed into the
definition of $M \to M \exp(2 \pi / \lambda_R) \equiv M$.

Our strategy is to study the energy of configurations of DNA strands that capture as much as possible
the symmetry of arrangements encountered in real cases \cite{Hud}. Hence, after having learned on the two-strand interaction,
we focus on many-body interactions where the strands are sitting at the sites of hexagonal lattices\footnote{We also present here
results for four DNA strands sitting at the vertices of a rhombus.} (see Fig.~\ref{steps} for the
case of 19 strands) and perform a careful analysis.

\begin{figure}
 \centering
  \includegraphics[height=.3\textheight]{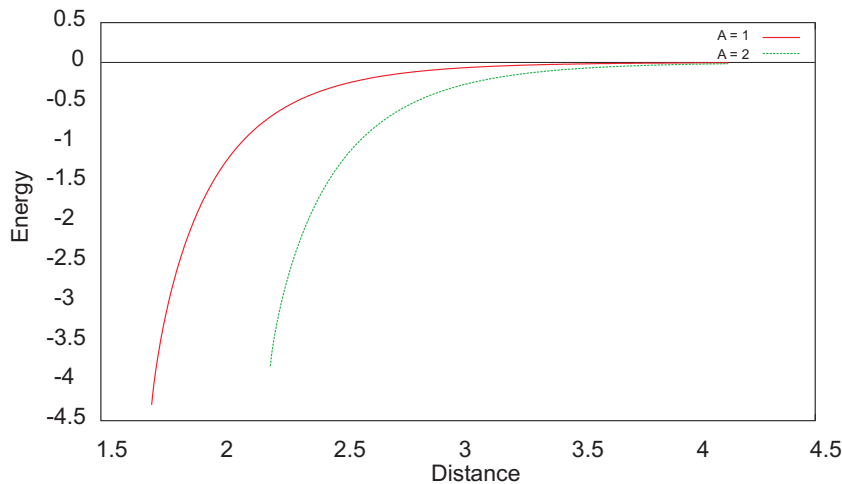}
  \caption{Interaction energy for the hexagonal lattice 19-strand configuration of Fig.~\ref{steps} with $a=1$ (upper curve) and $a=2$ (lower curve). The units are  $1 / \mu \sim {\cal O} (10)$\AA for the distances and ${\cal O} (10^{-1}) - {\cal O} (1)$~eV for the energy.}
\label{Two_19_String_Cases_Energy}
\end{figure}

To render the expression (\ref{Elndet}) suitable for such a study we first numerically perform
the integral over $E$ and then plot the resulting expression ${\cal E} (x)$ where
the relative distances $l_{i j} = c_{i j } x$ are expressed in terms of the basic lattice distance
$x$, the numerical coefficients $c_{i j}$ take the symmetry of the given arrangement into account, the distances are measured in units of $\mu^{-1} \sim {\cal O} (10)$ \AA, the other scale $M$ is constrained to be positive and less than $\mu (= 1)$ and we write it as $0 < M = e^{-1/a} < 1$. In this fashion the range of $\cal E$ scales with $a$ and we present here results for $a = 1$ and $a = 2$. To estimate the order of magnitude of the energy in the range of non-vanishing interaction we write the integral in Eq.(\ref{Elndet}) in a dimensionless fashion introducing the scale $\mu$. This way we obtain that the energy scale factor in front is $\hbar c \mu \simeq 6 \times 10^2$~eV, where $c$ is taken to be the speed of light in water, $c \simeq 2 \times 10^8 m s^{-1}$. Considering that the numerical values of the integrals we obtain are ${\cal O}(10^{-3})$, the estimate of the magnitude of the effect in the range is ${\cal E} \sim {\cal O} (10^{-1}) - {\cal O} (1) \; {\rm eV}$, that is between one and two orders of magnitude stronger than the thermal energy that at room temperature is $k_B T \simeq 2 \times 10^{-2}$~eV. These estimates clearly indicate that the quantum relativistic effect we are considering is important for explaining the attraction of DNA strands.

The two-strand interaction energy is shown in Fig.~\ref{2_String_Energy}. It is clearly attractive and finite-range. Similar attractive behaviors for the two-body interaction have been found in various models \cite{kornyshev1999, groenbech}. What we observe here is that in those models it is not clear why the interaction still needs to be attractive for more than two strands and why the aggregates are stable. For the Casimir energy we are considering here this is indeed the case, since this attraction mechanism does not suffer of any frustration.

\begin{figure}
 \centering
  \includegraphics[height=.3\textheight]{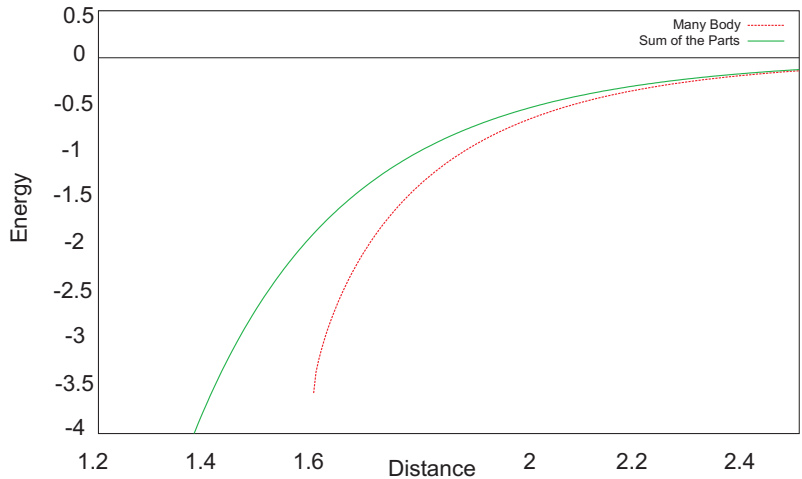}
  \caption{The lower curve is the interaction energy for the four-strand-rhombic configuration (many-body). The upper curve is what is obtained
  by summing-up the 6 two-body interactions. In both cases $a =2$. The units are  $1 / \mu \sim {\cal O} (10)$\AA~for the distances and ${\cal O} (10^{-1}) - {\cal O} (1)$~eV for the energy.}
\label{4_Rhombus_Many_Body_vs_Sum_of_Parts}
\end{figure}

To establish whether the magnitude and range of this attractive energy is indeed relevant for the case of DNA aggregates we need to move to the many-body case. We have computed the energy for several interacting strands having various configurations. We present in Fig.\ref{Two_19_String_Cases_Energy} the results for 19 strands arranged as in Fig.~\ref{steps}. More configurations are discussed in \cite{iss}. Comparing these plots with that of the two strands interaction we clearly see that the attraction becomes stronger and acts on a larger range when the number of strands increases. The range of attraction can be adjusted by fixing $a$ to fit the typical distances reached within the aggregates, that for the hexagonally packed toroidal condensates ranges between~\cite{Hud} 18~\AA~and 28~\AA, values clearly compatible with the range we obtain here.

\begin{figure}
 \centering
  \includegraphics[height=.3\textheight]{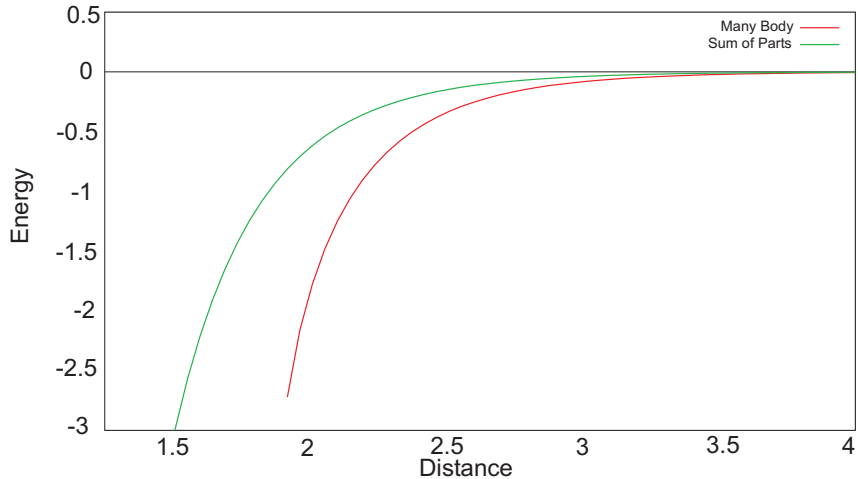}
  \caption{The lower curve is the interaction energy for the 7-strand configuration, i.e. for 7 strands sitting at the vertices and at the certer
  of a regular hexagon (many-body). The upper curve is what is obtained by summing-up the 21 two-body interactions. In both cases $a =2$. The units are  $1 / \mu \sim {\cal O} (10)$\AA~for the distances and ${\cal O} (10^{-1}) - {\cal O} (1)$~eV for the energy.}
\label{2K0_7_String_Many_Body_vs_Sum_of_Parts}
\end{figure}

In real cases it is always several DNA strands that interact, the two strands being only an idealization. Thus the fact that for 19 strands we find that (for $a=2$) the range of the force is in agreement with the typical values reported for DNA aggregates \cite{Hud} we take it as an indication of the validity of our hypothesis that the quantum Casimir energy holds together the aggregates. Furthermore, this force is many-body in nature and the many-body effects are big, another reason for taking the two-body interaction only as an indication of the real phenomenon. That the many body effects are strong we proved in our numerical calculations where we compared the $N$-body energy of Eq.~(\ref{Elndet}) with that obtained by summing up $(N/2)(N-1)$  two-body interactions. The results for four and seven strands are shown in Fig.~\ref{4_Rhombus_Many_Body_vs_Sum_of_Parts} and Fig.~\ref{2K0_7_String_Many_Body_vs_Sum_of_Parts}, respectively, and they indicate that the effect grows with $N$.

The singularity in the energy at a value $\bar{x}$ such that the determinant function becomes zero, is the indication of the limit of validity of our approximations. $\bar{x}$ can be evaluated for the various cases by plotting the determinant \cite{iss}. That singularity means that if only the Casimir force were present the strands would collapse to zero separation, an instance that does not occur in the real case because of nonlinear corrections to the Casimir force itself at every short distance and because of other forces (not considered here) such as the electrostatic force that for more than two strands will give a net repulsive effect. Another important factor at such short distances is of course the finite size of DNA strands that have a transverse length (radius of the cylinder) of $10$\AA.

The main result we present here is probably the demonstration that {\it quantum relativistic} effects can be
responsible for the collapse of DNA strands into aggregates (after the OM condensation has taken place) and
for holding them together into stable condensates. The time-dependent fluctuations of the electric field that
we studied are quantum in nature, propagate at the speed of light in the medium thus give rise to a Casimir force that
is attractive and short range for the two-body case and is many-body, the departures from the ``sum of two-bodies''
being important and growing with the number of strands. The magnitude and range of the interaction is such that it could
explain the formation and stability of DNA aggregates, as a preliminary comparison with reported data shows.
We focused our attention on the difficult problem of computing such interaction for several DNA strands and were able to overcome the
analytical challenges with numerical calculations performed for a variety of cases, most of which with hexagonal symmetry
of arrangement, the typical situation reported in experiments. Although our model is a simple one and to have the full picture it needs to be completed (by introducing other forces such as the hydrophobic, hydrophilic and electrostatic forces, and finite size effects, as well as by considering nonlinear corrections to the Casimir effect itself at very short distances $\bar x$) in the light of the results above listed we suggest that the Casimr vacuum energy can be the ``glue'' that holds together the DNA strands as aggregates.

{\bf Acknowledgments} A.I. has been supported in part by the Department of Mathematics
and Informatics of Salerno University. Siddhartha S. acknowledges
the kind hospitality of the Institute for Particle and Nuclear
Physics of Charles University of Prague.

\end{document}